\documentclass[12pt]{article}

\usepackage{scicite}

\usepackage{times,xspace,here,graphicx,amsmath,amssymb,deluxetable}
\usepackage{hyperref}
\usepackage[yyyymmdd,hhmmss]{datetime}
\usepackage{upgreek}
\usepackage{pdfpages}

\usepackage{color}

\newcommand{\um}{\ensuremath{\upmu\rm{m}}\xspace}
\newcommand{\kms}{km\,s\ensuremath{^{-1}}\xspace}
\newcommand{\Msol}{\ensuremath{\rm{M}_\odot}\xspace}
\newcommand{\Lsol}{\ensuremath{\rm{L}_\odot}\xspace}
\newcommand{\arc}{\ensuremath{''}\xspace}
\newcommand{\uJy}{\ensuremath{\mu\rm{Jy}}\xspace}
\newcommand{\Lz}{\ensuremath{\lambda_0}\xspace}
\newcommand{\lfir}{\ensuremath{L_{\rm{FIR}}}\xspace}
\newcommand{\lir}{\ensuremath{L_{\rm{IR}}}\xspace}
\newcommand{\gdr}{\ensuremath{\delta_{\rm{GDR}}}\xspace}
\newcommand{\Tdust}{\ensuremath{T_{\rm{dust}}}\xspace}
\newcommand{\Mdot}{\ensuremath{\dot M_{\rm{out}}}\xspace}
\newcommand{\Mout}{\ensuremath{M_{\rm{out}}}\xspace}
\newcommand{\Mgas}{\ensuremath{M_{\rm{gas}}}\xspace}
\newcommand{\reff}{\ensuremath{r_\mathrm{eff}}\xspace}
\newcommand{\vmax}{\ensuremath{v_{\rm{max}}}\xspace}
\newcommand{\alphaco}{\ensuremath{\alpha_{\rm{CO}}}\xspace}
\newcommand{\cii}{[C{\scriptsize II}]\xspace}

\newenvironment{sciabstract}{%
\begin{quote} \bf}
{\end{quote}}

\title{Fast Molecular Outflow from a Dusty Star-Forming Galaxy in the Early Universe}

\author{}

\date{}

\begin{document} 


\baselineskip24pt


\maketitle 

\begin{centering}\begin{large}
J.~S.~Spilker,$^{1,2,\ast}$
M.~Aravena,$^3$ 
M.~B\'ethermin,$^4$ 
S.~C.~Chapman,$^5$ 
C.-C.~Chen,$^6$
D.~J.~M.~Cunningham,$^{5,7}$
C.~De~Breuck,$^6$ 
C.~Dong,$^8$ 
A.~H.~Gonzalez,$^8$ 
C.~C.~Hayward,$^{9,10}$
Y.~D.~Hezaveh,$^{11}$ 
K.~C.~Litke,$^{2}$ 
J.~Ma,$^{12}$ 
M.~Malkan,$^{13}$, 
D.~P.~Marrone,$^{2}$ 
T.~B.~Miller,$^{5,14}$ 
W.~R.~Morningstar,$^{11}$ 
D.~Narayanan,$^8$ 
K.~A.~Phadke,$^{15}$
J.~Sreevani,$^{15}$ 
A.~A.~Stark,$^{10}$ 
J.~D.~Vieira,$^{15}$ 
A.~Wei\ss$^{16}$\\
\end{large}\end{centering}

\vspace{0.2in}
\begin{centering}\begin{normalsize}
$^{1}$Department of Astronomy, University of Texas at Austin, 2515 Speedway Stop C1400, Austin, TX 78712, USA\\
$^{2}$Steward Observatory, University of Arizona, 933 North Cherry Avenue, Tucson, AZ 85721, USA\\
$^{3}$N\'ucleo de Astronom\'{\i}a, Facultad de Ingenier\'{\i}a, Universidad Diego Portales, Av. Ej\'ercito 441, Santiago, Chile\\
$^4$Aix Marseille Univ., Centre National de la Recherche Scientifique, Laboratoire d'Astrophysique de Marseille, Marseille, France\\
$^5$Department of Physics and Atmospheric Science, Dalhousie University, Halifax, Nova Scotia, Canada\\
$^6$European Southern Observatory, Karl Schwarzschild Stra\ss e 2, 85748 Garching, Germany\\
$^7$Department of Astronomy and Physics, Saint Mary's University, Halifax, Nova Scotia, Canada\\
$^8$Department of Astronomy, University of Florida, Bryant Space Sciences Center, Gainesville, FL 32611, USA\\
$^9$Center for Computational Astrophysics, Flatiron Institute, 162 Fifth Avenue, New York, NY 10010, USA\\
$^{10}$Harvard-Smithsonian Center for Astrophysics, 60 Garden Street, Cambridge, MA 02138, USA\\
$^{11}$Kavli Institute for Particle Astrophysics and Cosmology, Stanford University, Stanford, CA 94305, USA\\
$^{12}$Department of Physics and Astronomy, University of California, Irvine, CA 92697, USA\\
$^{13}$Department of Physics and Astronomy, University of California, Los Angeles, CA 90095, USA\\
$^{14}$Department of Astronomy, Yale University, 52 Hillhouse Avenue, New Haven, CT 06511, USA\\
$^{15}$Department of Astronomy, University of Illinois, 1002 West Green St., Urbana, IL 61801, USA\\
$^{16}$Max-Planck-Institut f\"{u}r Radioastronomie, Auf dem H\"{u}gel 69 D-53121 Bonn, Germany\\
$^\ast$Corresponding author. E-mail:  spilkerj@gmail.com.\\
\end{normalsize}\end{centering}

\renewcommand{\thefigure}{\arabic{figure}.}
\setcounter{figure}{0}

\begin{sciabstract}

Galaxies grow inefficiently, with only a few percent of the available gas converted into stars each free-fall time. Feedback processes, such as outflowing winds driven by radiation pressure, supernovae or supermassive black hole accretion, can act to halt star formation if they heat or expel the gas supply. We report a molecular outflow launched from a dust-rich star-forming galaxy at redshift 5.3, one billion years after the Big Bang. The outflow reaches velocities up to 800\,\kms relative to the galaxy, is resolved into multiple clumps, and carries mass at a rate within a factor of two of the star formation rate. Our results show that molecular outflows can remove a large fraction of the gas available for star formation from galaxies at high redshift.

\end{sciabstract}

The formation of realistic populations of galaxies requires one or more forms of self-regulating feedback to suppress the conversion of gas into stars. Cosmological simulations invoke various mechanisms to regulate star formation, in the form of energy deposition and wind launching linked to supermassive black hole activity, supernovae, and/or radiation pressure from massive stars. The strength and scalings of these processes play critical roles in the evolution of galaxies by regulating the growth of stellar mass relative to the dark matter halo, connecting the properties of central black holes to their host galaxies, and enriching the circum-galactic medium with heavy elements \cite{murray05,hopkins08,kormendy13,ceverino16}.

Outflowing winds of gas are ubiquitous in nearby galaxies. The gas in outflows spans many orders of magnitude in temperature and density \cite{veilleux05,leroy15,schneider17}, and as such different components of the winds are observable from X-ray to radio wavelengths. Observing  winds in the distant Universe is difficult: not only are the spectral features faint, but outflow tracers observed in emission may be less reliable due to the ongoing processes of galaxy assembly \cite{narayanan15}. The constrained geometry of absorption lines provide signatures of inflowing and outflowing material, but have thus far eluded detection.

Passive galaxies with stellar masses $\sim10^{11}$\,\Msol, low star formation rates (SFRs, $\lesssim 10$\,\Msol\,yr$^{-1}$) and stellar ages $\sim$0.8\,Gyr were already in place when the Universe was 2\,Gyr old, implying these galaxies formed stars at rates of hundreds of \Msol\,yr$^{-1}$ before $z=5$ (where $z$ is the redshift) \cite{straatman14}. Galaxies with such high SFR are extremely rare in rest-ultraviolet surveys, implying that such galaxies are incapable of becoming sufficiently massive by $z\sim4$ to reproduce the observed passive population. Instead, this suggests a connection between early passive galaxies and high-redshift dusty, star-forming galaxies (DSFGs) observed at far-infrared wavelengths\cite{blain04,tacconi08,glazebrook17}.  With a redshift distribution that includes a significant number of objects at $z>4$ \cite{vieira13,strandet16}, DSFGs represent a plausible progenitor population of the earliest passive galaxies. If this evolutionary connection is correct, many DSFGs should show signs of the feedback process(es) acting to suppress their rapid star formation.

We present observational evidence of a massive molecular wind being launched from SPT$-$S J231921$-$5557.9 (SPT2319$-$55 hereafter), a DSFG observed when the Universe was only one billion years old. SPT2319$-$55 was discovered in the 2500deg$^2$ South Pole Telescope survey \cite{mocanu13} on the basis of its thermal dust emission. Earlier observations of this source from the Atacama Large Millimeter/submillimeter Array (ALMA) determined its redshift to be $z_{\rm{source}} = 5.293$ \cite{strandet16} and showed that it is gravitationally lensed by an intervening foreground galaxy \cite{spilker16}. As is typical for these objects, SPT2319$-$55 is both gas-rich, containing $\sim1.2\times 10^{10}$\,\Msol of molecular gas, $\sim1.2\times10^8$\,\Msol of dust, and is forming stars very rapidly, SFR\,$\sim790$\,\Msol\,yr$^{-1}$ \cite{scimethods} (these values account for the lensing magnification).

We used ALMA to observe the rest-frame 119\,\um ground-state doublet transition of the hydroxyl molecule, OH, and the thermal dust emission at this wavelength \cite{scimethods}. This transition is a good tracer of gas flows in nearby galaxies \cite{sturm11,veilleux13}. The ALMA observations reach a spatial resolution of $0.25'' \times 0.4''$, and resolve the lensed images of SPT2319$-$55 (Figure~1). We detect a molecular outflow from SPT2319$-$55, seen in blueshifted absorption against the bright dust continuum emission, a signature of outflowing molecular material (Figure~1). 

We fit the spectrum in Figure 1 with the sum of two velocity components, each consisting of two equal-amplitude Gaussian profiles separated by 520\,\kms, the separation of the two components of the OH doublet \cite{scimethods}. As is common practice for low-redshift observations of OH, we assign the higher-frequency component of the doublet to the systemic velocity of the galaxy. Because the redshift of this source is known to better than 50\,\kms from other observations \cite{strandet16}, we fix the velocity offset of one pair of Gaussian profiles to the systemic velocity. This component represents absorption due to gas from within the galaxy, with a fitted full-width-at-half-maximum (FWHM) linewidth of $330\pm80$\,\kms. We allow a velocity offset for the second pair of Gaussian profiles; this component represents the blueshifted molecular outflow. We find that this second component is blueshifted relative to the galaxy by $440\pm50$\,\kms, and derive a maximum velocity of $\sim$800\,\kms \cite{scimethods}. 

The ALMA observations spatially resolve and clearly detect the dust continuum even in relatively narrow velocity channels. Because SPT2319$-$55 is gravitationally lensed by a foreground galaxy, we determine its intrinsic structure using a lens modeling technique that represents the galaxy as an array of pixels \cite{scimethods,hezaveh16}. In addition to the line-free continuum emission, we also reconstruct the OH absorption components using velocity ranges relative to the higher-frequency OH transition of -700 to -200\,\kms for the wind component and +300 to +700\,\kms for the internal component; the latter velocity range corresponds to velocities -220 to +180\,\kms relative to the lower-frequency transition, and traces gas within SPT2319$-$55. These velocity ranges fairly cleanly separate the wind and internal absorption \cite{scimethods}. 

The doubly-imaged continuum emission in Figure~1 is consistent with the lensing of a single background galaxy, with a well-determined extent, by a single foreground lens. The reconstructions of the dust continuum and each absorption component are shown in Figure~2. The continuum is dominated by a single bright region $\sim$1.2\,kiloparsecs (kpc) in diameter \cite{scimethods}, with flux density $S_{\mathrm{400GHz}} = $9.0\,mJy magnified by a factor $\mu = 5.8$. The internal absorption, arising from gas within the galaxy, is concentrated towards the center of the object. This is to be expected, as the continuum is brightest and gas column densities are highest towards the nuclear region, making the detection of absorption easier.

The geometry of the molecular outflow is more complex, and is not confined to the center of the galaxy. Instead, it is clustered into multiple clumps, separated from each other by a few hundred parsecs, corresponding to $\approx$2.5 FWHM resolution elements \cite{scimethods}. Because absorption is easier to detect against a strong continuum, we would expect a geometry similar to that of the internal absorption if the continuum were the limiting factor in the reconstruction. Tests using mock data also show that absorbing components weaker than the outflow are well recovered by our reconstructions \cite{scimethods}.

The overall covering fraction of the wind is high -- 80\% of pixels with continuum signal-to-noise $>8$ have significant wind absorption (although these pixels are not all independent; \cite{scimethods}). If the covering fraction along the line of sight we observe is the same as the rest of the source, this implies a total wind opening angle of $\sim0.8\times4\pi$ steradians. On the other hand, we can set a rough lower limit on the opening angle if we assume that we have observed the entirety of the outflow, and that the molecular material is destroyed by the time it reaches a few galaxy radii away, similar to local starburst galaxies \cite{leroy15}. If the maximum radius of the outflowing material is 2 galaxy radii, for example, then the solid angle of a sphere of this radius subtended by the source is $(\pi \reff^2)/(4\pi(2\reff)^2)\times4\pi$ steradians, with \reff the radius of the galaxy. With an 80\% covering fraction of the source along the line of sight, this implies a minimum opening angle of $0.2\pi$ steradians.

In low-redshift galaxies, the 119\,\um OH transitions are very optically thick \cite{gonzalezalfonso17}, and the absorption depth thus directly corresponds to the covering fraction of the continuum.  The difference between the 80\% covering fraction from our lens modeling and the peak wind depth ($\sim$15\%) mirrors the discrepancy between the typical absorption depths and high detection rate of winds in low-redshift objects studied in OH \cite{veilleux13,spoon13}. If the OH absorption in SPT2319$-$55 is also optically thick, we expect that the wind contains significant substructure on scales below our effective resolution limit, with most of the absorption arising from small, highly optically-thick clumps. Alternatively, because the outflow reconstruction spans a wide range of velocities, it is possible that the covering fraction is lower in narrower velocity ranges. In this case we also expect significant substructure on smaller scales, as the lower covering fractions in narrow velocity bins must still sum to the high overall covering fraction we have observed. On the other hand, if the OH excitation temperature is high, the OH molecules can substantially fill in the absorption profile with re-emitted 119\,\um photons, reducing the absorption strength while maintaining a high covering fraction without small-scale structure in the wind, although this implies high densities in the absorbing gas.

We estimate the mass outflow rate using a simple model for the outflow geometry \cite{scimethods}. The primary uncertainties in this estimate are the unknown optical depth of the OH 119\um doublet and the detailed geometry of the wind. In the limiting case of optically-thin absorption, we find a minimum mass outflow rate $\Mdot \gtrsim 60$\,\Msol\,yr$^{-1}$. A more likely outflow rate, based on the empirical correlation between OH equivalent width and \Mdot in low-redshift dusty galaxies \cite{scimethods}, is $\Mdot \sim 510$\,\Msol\,yr$^{-1}$. While highly uncertain \cite{scimethods}, this value is within a factor of two of the SFR of SPT2319$-$55, indicating that the wind is capable of depleting the molecular gas reservoir on a timescale similar to star formation itself. We compare the SFR and molecular outflow rates we derive for SPT2319$-$55 under both assumptions in Figure 3, along with molecular outflow rates for low-redshift objects \cite{cicone14,gonzalezalfonso17}. 

The mass loading factor of the wind, \Mdot/SFR$\sim$0.7, is high even accounting for molecular material alone; the inclusion of unobserved wind phases, namely neutral atomic and ionized gas, would increase this value still further. High mass loading factors are expected from simulations of galaxies \cite{muratov15,hayward17}, which require strong feedback to prevent the overproduction of stars. On the other hand, both the origin of the molecules in the SPT2319$-$55 wind (whether entrained in hotter material or formed in situ) as well as the driving source are unclear. Current data place an upper limit of 30\% on the contribution of an active galactic nucleus to the total luminosity of SPT2319$-$55, but cannot rule out nuclear activity at a lower level \cite{scimethods}. 

Despite the high overall outflow rate, it is likely that only a small fraction of the wind mass is sufficiently fast-moving to escape the gravitational potential of the galaxy. Given the molecular gas mass of SPT2319$-$55 and assuming a typical gas fraction for DSFGs at this redshift \cite{aravena16} and using the size from the source reconstruction yields an escape velocity from the galaxy of $\sim$650\,\kms.  The fraction of the absorption at speeds greater than this value indicates that $\sim$10\% of the outflowing material will escape SPT2319$-$55, under the same assumptions as the outflow as a whole. This escape fraction is within the range observed in local galaxies \cite{alatalo15a}. The remainder of the material, therefore, is destined to remain within the galaxy's dark matter halo.

Our results show that self-regulating feedback is acting to disrupt and remove the molecular gas in SPT2319$-$55, and will likely suppress the rapid star formation in this galaxy in $\lesssim$100\,Myr. Whether this is sufficient to quench the star formation on a more permanent basis is less clear. A large fraction of the wind material is likely to remain within the galaxy, capable of fueling future star formation at later times. On the other hand, SPT2319$-$55 likely resides in a dark matter halo of mass $\sim10^{12}$\,\Msol \cite{behroozi13}, sufficiently massive to shock infalling gas to high temperatures and prevent its accretion \cite{dekel06}. Other forms of feedback, such as from an active galactic nucleus, may also act to maintain the low SFR after the initial suppression \cite{fabian12}. Molecular outflows like the one we have observed are likely an important step in the suppression of star formation and the emergence of early passive galaxies.

\clearpage

\begin{figure*}[htb]
\centering
\includegraphics[width=\textwidth]{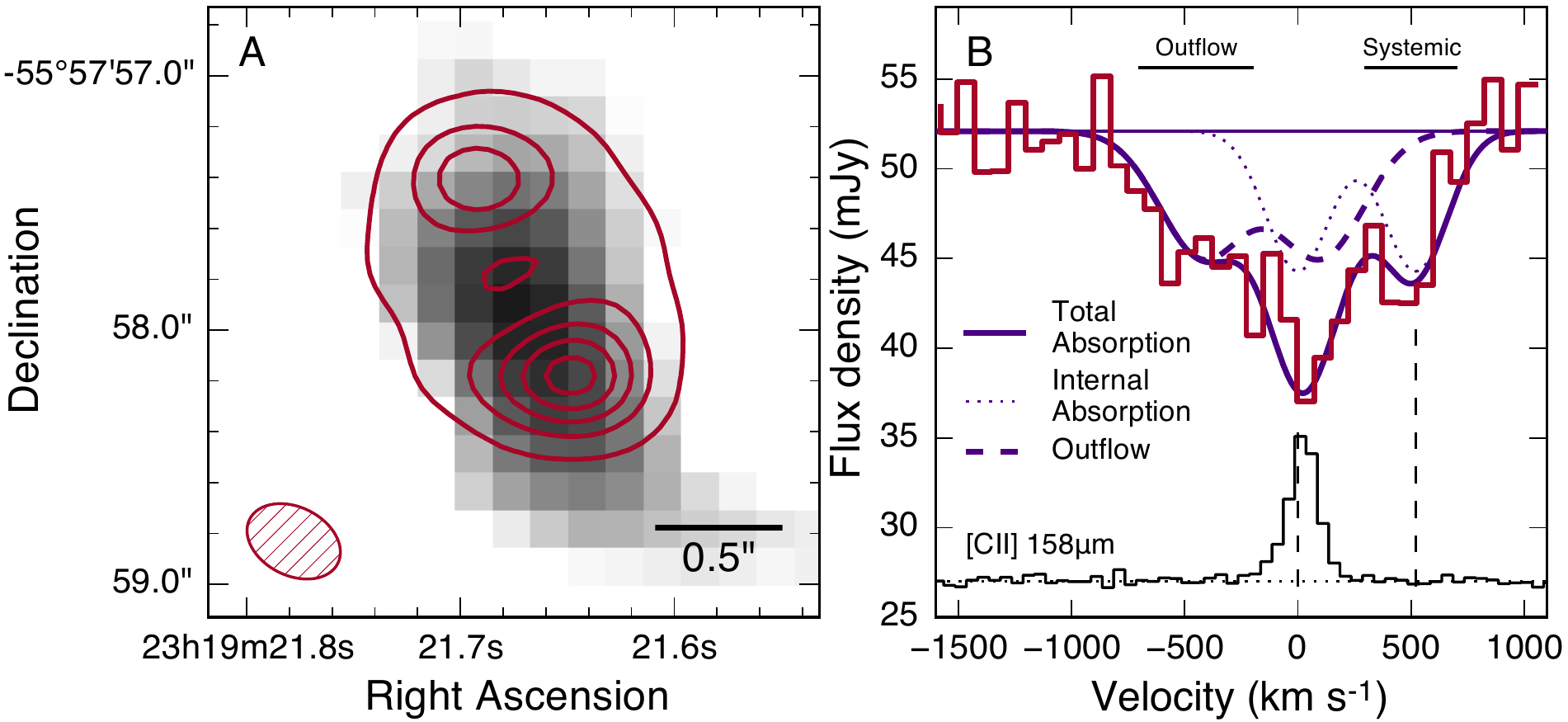}
\caption{
\textbf{ALMA 400\,GHz continuum image and OH spectrum of SPT2319$-$55.}
\textbf{A.} The ALMA continuum data (red contours) overlaid on a 2.2\,\um image of the foreground lens galaxy. Contours are drawn at 10, 30, 50, 70, and 90\% of the peak value; the data reach a peak signal-to-noise of $\sim$140. The synthesized beam is shown with a hatched ellipse at lower left.
\textbf{B.} The integrated apparent (not corrected for lensing magnification) OH 119\,\um spectrum of SPT2319$-$55, with velocity scale relative to the higher-frequency component of the OH doublet. The rest velocities of the two doublet components are shown with vertical dashed lines, and we show an ALMA \cii spectrum of this source \cite{scimethods} as an indication of the linewidth of this galaxy due to internal gas motions (with arbitrary vertical normalization). We fit the OH 119\,\um spectrum as described in the text; the navy dotted line shows the component due to gas within the galaxy, the navy dashed line the blueshifted outflow component, and the solid line the total absorption profile. We mark the velocity ranges for which we create lens models with horizontal bars above the continuum.
}
\label{fig:outflowdata}
\end{figure*}

\begin{figure*}[htb]
\centering
\includegraphics[width=\textwidth]{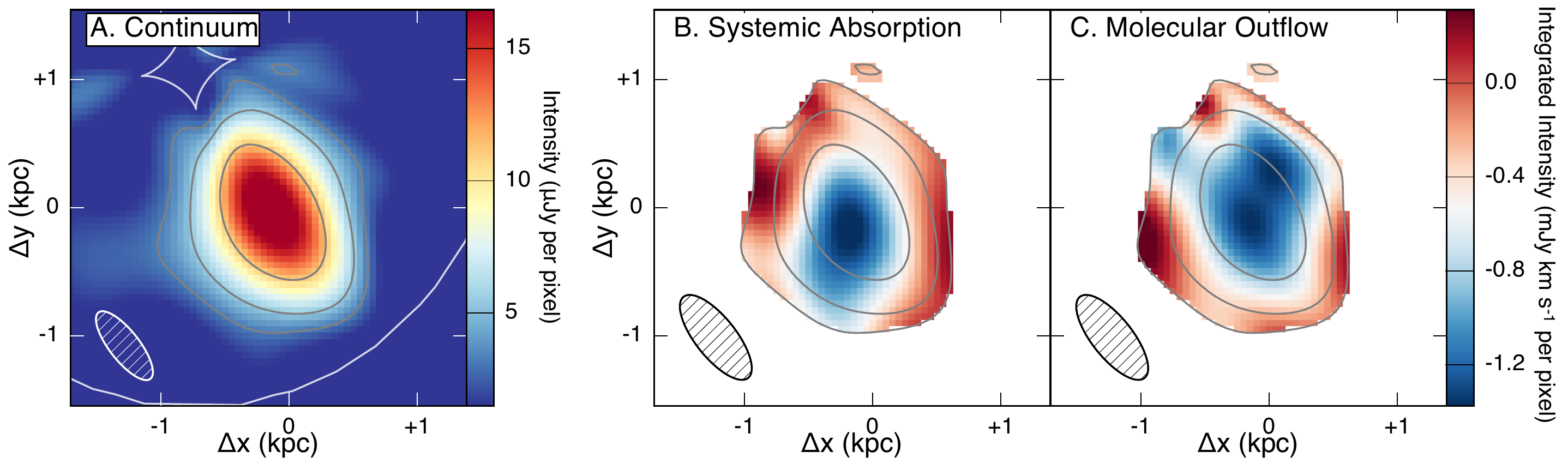}
\caption
{
\textbf{Lensing reconstruction of the continuum and OH absorption in SPT2319$-$55.}
\textbf{A.} Source-plane structure of the rest-frame 119\,\um continuum emission after modeling the effects of gravitational lensing. Contours are shown at signal-to-noise ratios of 8, 16, and 32; these contours are repeated in subsequent panels. The ellipse at lower left shows the effective resolution of the reconstruction \cite{scimethods}. The lensing caustics (lines of theoretically infinite magnification) are also shown. Coordinates are offset relative to the ALMA phase center.
\textbf{B.} Reconstruction of the absorption due to gas internal to SPT2319$-$55. 
\textbf{C.} Reconstruction of the molecular outflow.
Because the reconstruction of the absorption requires the presence of continuum emission, in panels B and C we mask pixels with continuum signal-to-noise $<8$. We separate the internal and outflow absorption using the velocity ranges labeled in Figure~1. Ellipses at lower left indicate the effective spatial resolution of the absorption reconstructions \cite{scimethods}.
}
\label{fig:srcrec}
\end{figure*}

\begin{figure*}[htb]
\centering
\includegraphics[width=0.6\textwidth]{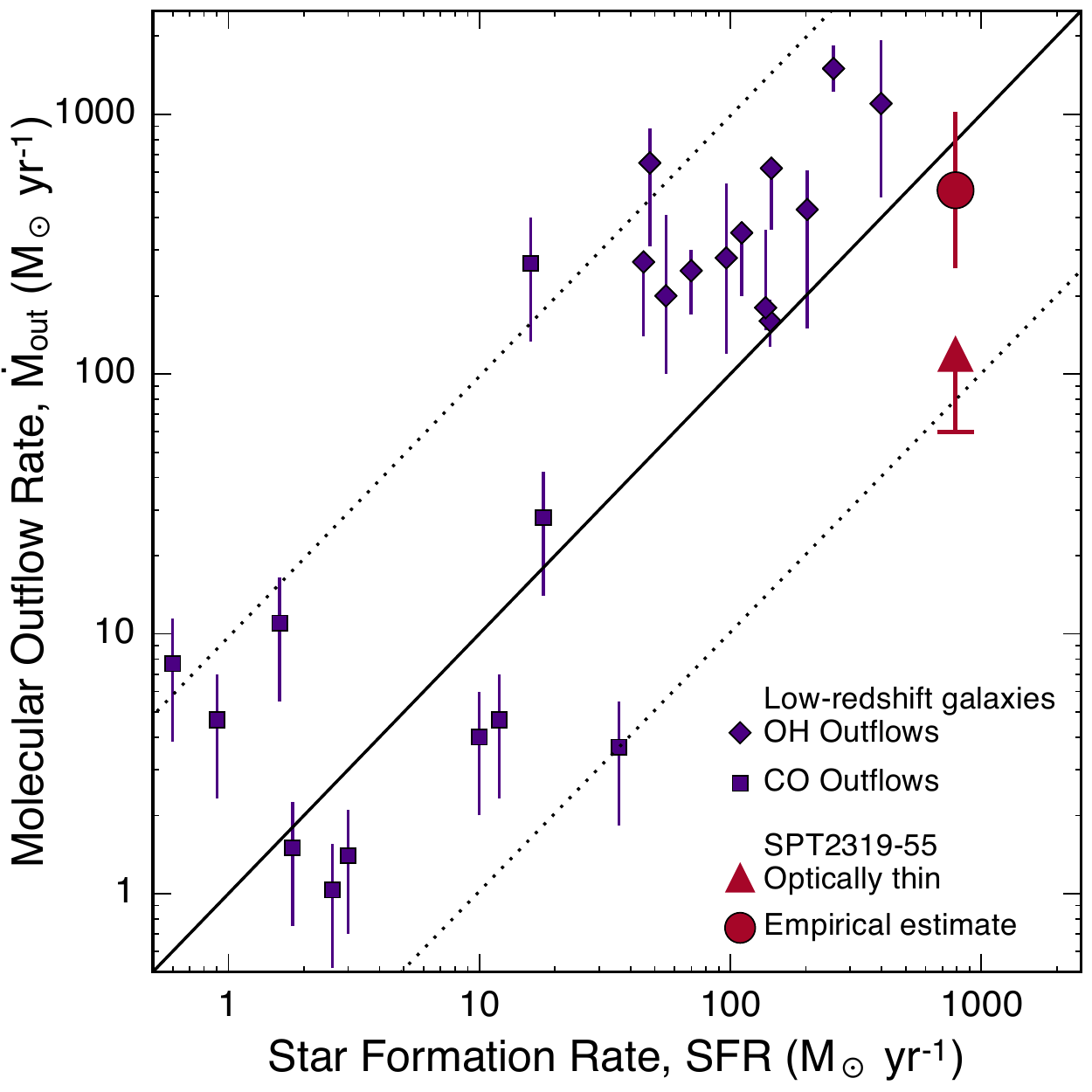}
\caption{
\textbf{Molecular outflow rates as a function of the star formation rate.} Outflow rates are from both OH and CO \cite{cicone14,gonzalezalfonso17}. The solid line shows the one-to-one relation, with dotted lines a factor of ten above and below. The low-redshift samples include both sources dominated by star formation and sources with active galactic nuclei.
}
\label{fig:outflowmdot}
\end{figure*}

\clearpage

\bibliographystyle{Science}
\bibliography{outflow_main.bbl}

\section*{Acknowledgements}
An allocation of computer time from UA High Performance Computing at the University of Arizona is gratefully acknowledged, and the authors acknowledge the Texas Advanced Computing Center (TACC) at The University of Texas at Austin for providing HPC resources that have contributed to the results reported here.

\section*{Funding}
ALMA is a partnership of ESO (representing its member states), NSF (USA) and NINS (Japan), together with NRC (Canada) and NSC and ASIAA (Taiwan), in cooperation with the Republic of Chile. The Joint ALMA Observatory is operated by ESO, AUI/NRAO and NAOJ.
The SPT is supported by the National Science Foundation through grant PLR-1248097, with partial support through PHY-1125897, the Kavli Foundation and the Gordon and Betty Moore Foundation grant GBMF 947. 
The Australia Telescope Compact Array is part of the Australia Telescope National Facility which is funded by the Australian Government for operation as a National Facility managed by CSIRO.
J.S.S. thanks the McDonald Observatory at the University of Texas at Austin for support through a Smith Fellowship.
J.S.S., K.C.L., D.P.M., S.J., and J.D.V. acknowledge support from the U.S. National Science Foundation under grant AST-1312950; K.C.L and D.P.M also acknowledge support under AST-1715213 and S.J. and J.D.V also acknowledge support under AST-1716127.
C.C.H acknowledges support from The Flatiron Institute, which is supported by the Simons Foundation.
Y.D.H. is a Hubble fellow.

\section*{Author Contributions}
J.S.S. proposed the ALMA OH observations, performed the data reduction and lensing analysis, and wrote the manuscript. D.P.M. and K.C.L. led the observations of \cii shown in Figure~1. M.A. led the CO observations used to calculate the molecular gas mass. K.A.P. and J.D.V. performed modeling of the source spectral energy distribution. All authors discussed the results and provided comments on the figures and text, and are ordered alphabetically after J.S.S.

\section*{Competing Interests}
The authors declare no competing interests.

\section*{Data and Materials Availability}
This paper makes use of the following ALMA data: ADS/JAO.ALMA\#2016.1.00089.S and ADS/JAO.ALMA\#2016.1.01499.S, archived at \url{https://almascience.nrao.edu/alma-data/archive}. Pixelated reconstructions were performed using a lens modeling tool developed by a subset of the authors and additional non-authors. Developers of this tool include authors Y.D. Hezaveh and W.R. Morningstar, and non-authors N. Dalal, G. Holder, and P. Marshall. Reduced data products, lens tool data files, and lensing reconstructions are provided at \url{https://github.com/jspilker/s18_outflow}.

\section*{Supplementary materials}
Materials and Methods\\
Figs. S1 to S8\\
Tables S1 and S2\\
References \textit{(31--61)}




\clearpage
\renewcommand{\thefigure}{S\arabic{figure}}
\setcounter{figure}{0}
\setcounter{page}{0}
\renewcommand{\thetable}{S\arabic{table}}
\renewcommand{\theequation}{S\arabic{equation}}

\pagenumbering{gobble}
\includepdf[pages=-,pagecommand={}]{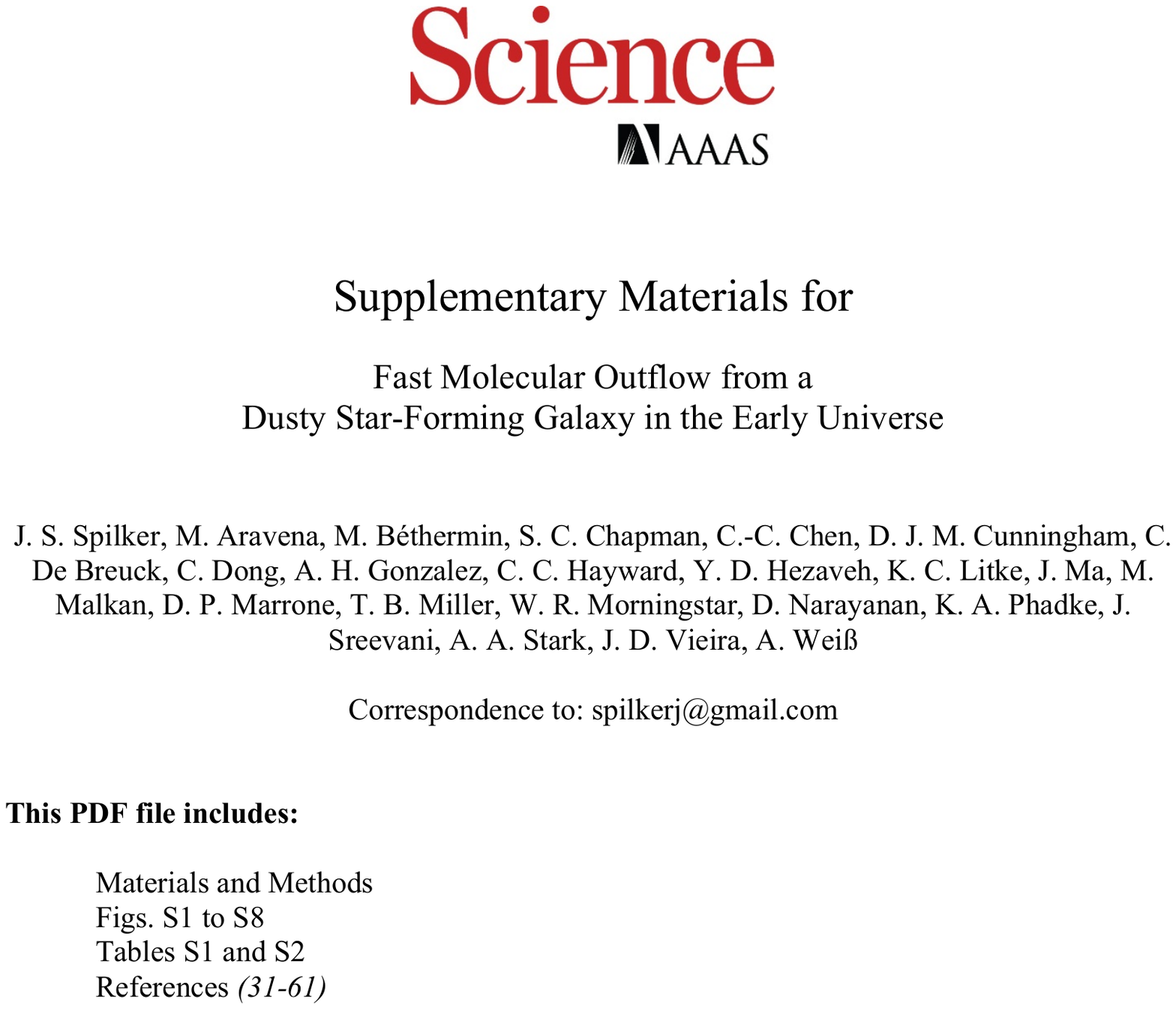}
\pagenumbering{arabic}

\section*{Materials and Methods}

\section{Target Selection and Observations}

\subsection{Target Selection}

SPT2319$-$55 was observed as part of ongoing ALMA program 2016.1.00089.S (PI: J. Spilker). This program aims to detect and spatially resolve massive molecular outflows using the ground-state OH $^2\Pi_{3/2}$ $J=5/2-3/2$ $\Lambda$ doublet in a subsample of DSFGs drawn from the South Pole Telescope (SPT) survey \cite{vieira10,mocanu13}. The two components of the doublet, at rest-frame 2509.9 and 2514.3\,GHz, are separated by $\sim$520\,\kms. Each exhibits additional hyperfine structure, but the individual hyperfine transitions are separated by $<$5\,\kms and are not resolved by our data or sufficiently well-separated to cause spectral confusion in our subsequent analysis. We use the rest frequencies determined with spectroscopic data from the NASA Jet Propulsion Laboratory line catalog \cite{pickett98}.

The primary selection criterion for our observations of the OH doublet was that the source redshift place the OH lines at frequencies of relatively good atmospheric transmission. A sample was selected from the 39 SPT-discovered DSFGs with spectroscopic redshifts determined using ALMA observations of CO and [C~\textsc{I}] \cite{weiss13,strandet16}. Observations of local galaxies showed that outflow velocities up to 1000\,\kms are possible \cite{veilleux13}, and so we restricted the sample to objects at $z>4$ to ensure that the fractional bandwidth of ALMA would be sufficient to cover both doublet transitions as well as the blueshifted outflow. Finally, we restricted the sample to objects with good lens models from ALMA 870\,\um observations \cite{spilker16}, and chose objects to span a wide range in SFR and SFR surface density, resulting in a final sample of 7 objects. 

\begin{figure*}[htb]
\centering
\includegraphics[width=0.5\textwidth]{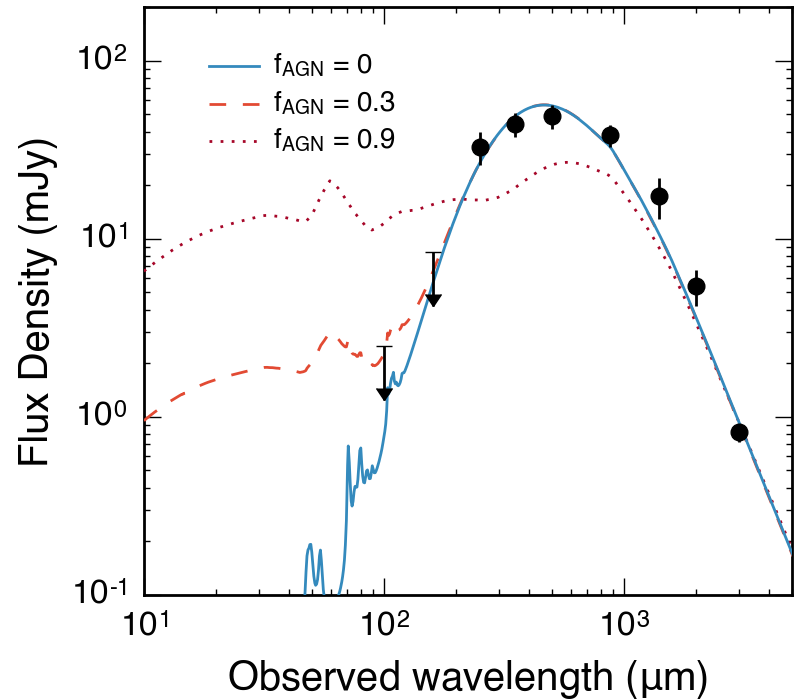}
\caption{
\textbf{Far-infrared and millimeter photometry of SPT2319$-$55 and fits to the spectral energy distribution.} The model fits were generated with the \textsc{CIGALE} modeling code. We show three different models, assuming $f_\mathrm{AGN} = 0$, 0.3, and 0.9. The non-detections at 100 and 160\,\um (plotted as 1$\sigma$ upper limits) motivate our conclusion that $f_\mathrm{AGN} < 0.3$ (also 1$\sigma$).
}
\label{fig:cigaleseds}
\end{figure*}

Some basic properties of SPT2319$-$55 are given in Table~\ref{tab:outflowspt2319data}. The infrared (IR; 8--1000\,\um) and far-infrared (FIR; 40--120\,\um) luminosities were determined by fitting to the FIR and submillimeter photometry, including non-detections by the \textit{Herschel Space Observatory} at 100 and 160\,\um \cite{strandet16}. Both simple (modified blackbody with mid-infrared powerlaw) and more complex models (using, e.g., \textsc{CIGALE}; \cite{burgarella05,noll09}) yield consistent results. The latter method yields an upper limit on the fractional AGN contribution to the bolometric luminosity of $f_\mathrm{AGN} < 30$\% (1$\sigma$). This limit is driven mostly by the PACS non-detections in the rest-frame mid-infrared, which probes hot dust near the nuclear torus. If the galaxy is optically thick at these wavelengths, however, a deeply embedded AGN may still be present in this system. We show the FIR and millimeter photometry in Figure~\ref{fig:cigaleseds}, with three different fits assuming $f_\mathrm{AGN} = 0$, 0.3, and 0.9.  The SFR was determined simply as SFR $= 10^{-10}\lir$, with SFR in units of \Msol\,yr$^{-1}$ and \lir in \Lsol. In hydrodynamical simulations of galaxies similar to SPT2319$-$55, the SFR often varies rapidly as bursts of star formation are countered by powerful feedback processes \cite{narayanan15}. However, SFRs observationally derived from the IR luminosity are sensitive to star formation on timescales as long as $\sim100$\,Myr \cite{kennicutt12}, so the SFR we measure for SPT2319$-$55 already likely averages over some of these rapid fluctuations.

The molecular gas mass was determined through observations of the CO(2--1) line using the Australia Telescope Compact Array (ATCA). The CO luminosity was converted to a molecular gas mass using the same assumptions as ref. \cite{aravena16}, namely, a CO(2--1)/CO(1--0) line ratio of 0.9 in temperature units \cite{bothwell13,spilker14}, and CO--H$_2$ conversion factor $\alphaco=0.8$ \cite{spilker15}. The dust mass was calculated from the long-wavelength photometry, sensitive to the Rayleigh-Jeans tail of the dust emission, under standard assumptions about the dust emissivity \cite{spilker15}. Comparison of the molecular gas and dust masses indicates a gas-to-dust mass ratio $\gdr\sim100$, typical of metal-rich star-forming galaxies \cite{sandstrom13}.

ALMA has also observed \cii emission in this source in project 2016.1.01499.S, which we use to produce the comparison spectrum in Figure~1. These data will be analyzed in detail in future work, but for the purposes of this work we extract the \cii spectrum in identical fashion to the OH spectrum as detailed below. We use this spectrum solely to compare the shape of the line profile to the OH line profile, so for the purposes of Figure~1 the spectrum has been rescaled and offset from zero continuum level for easier comparison.

\subsection{ALMA Observations and Data Analysis}

ALMA observed SPT2319$-$55 on 2016 November 18 using the Band 8 receivers \cite{satou08} tuned to a local oscillator frequency of 405.781\,GHz. Forty antennas were available at the time of observation, with baselines spanning 15--920\,m. Weather conditions were excellent, with precipitable water vapor of 0.5\,mm and typical system temperatures of 250\,K. The observations lasted 75\,min in total, with 30\,min spent on-source. Quasar PKS J2258-2758 served as the bandpass calibrator, while quasar PKS J2357-5311 was used both to calibrate the absolute flux scale and the time-varying antenna gains. Finally, the quasar PKS J2336-5236 was observed three times over the course of the observations to test the quality of the calibration and the absolute astrometric accuracy.

The ALMA correlator was configured to observe the OH doublet in the lower sideband, delivering two 1.875\,GHz basebands with 3.9\,MHz channels. These two basebands were placed to provide continuous frequency coverage, with a small overlap in frequency to insure against decreased sensitivity at the edges of each baseband. In total, these basebands yielded 3.63\,GHz of continuous frequency coverage, corresponding to $\sim$2700\,\kms of velocity coverage. The upper sideband was configured with two 2\,GHz basebands and 15.625\,MHz channels for continuum observations. 

The data were reduced using the standard ALMA Cycle 4 pipeline in \textsc{casa} \cite{mcmullin07}, and the calibration appears reasonable. From the observations of the test source PKS J2336-5236, we estimate the astrometric accuracy of the observations to be $\sim$0.07\arc. To estimate the noise in the visibility data, we difference visibilities of successive integrations from the same baseline, polarization, and baseband [see refs. \cite{hezaveh13,spilker16}].  As is typical for such high-frequency observations, the sensitivity of the data is not uniform over the observed bandwidth due to the frequency-dependent atmospheric transmission. Our data in particular are affected by telluric ozone lines at 400.077 and 401.218\,GHz, which increase the noise  by $\sim$20 and 30\%, respectively, within a few tens of MHz ($\lesssim50$\,\kms) of these frequencies. These telluric lines are sufficiently narrow with respect to our subsequent modeling to have no appreciable impact on our analysis. Finally, in order to decrease the number of visibilities for our lens modeling procedure, we average the data in time in intervals of up to 3\,min, unless doing so would cause individual visibilities in the average to have Fourier-plane $uv$ coordinates separated by $>$10\,m. This $uv$-plane maximum bin size would cause a source 3\arc from the phase center to suffer a decrease in amplitude by $<$1\% \cite{cotton09}, and thus introduces no serious de-correlation or loss of signal.

\subsection{Imaging and Spectral Analysis}

The continuum data reach a sensitivity of 105\,\uJy per beam and a beam size of 0.25$\arc\times0.4\arc$  for a naturally-weighted image, which maximizes sensitivity at the expense of spatial resolution. The source is detected at a peak signal-to-noise of $\sim$140. To search for OH absorption or emission, we  imaged the lower sideband data, averaging together 25 channels to create an image cube with a spectral resolution of $\sim$75\,\kms. This cube reaches a typical sensitivity of 45--55\,mJy\,\kms, and typically detects the dust continuum at a peak signal-to-noise of $\sim$30 in each channel even in the deepest portion of the OH absorption lines. To extract a spectrum of the source from this cube, we perform aperture photometry over the region within which the continuum is detected at $>3\sigma$.  We have verified that this procedure accurately recovers the total source flux using additional images created by applying a taper in the $uv$ plane.  Because the two brightest peaks in the continuum emission in Figure~1 are gravitationally-lensed images of the same regions of the galaxy, we have also verified that the OH absorption spectra extracted from each peak separately are consistent. The OH spectrum of this source is complex, but clearly shows both systemic and significantly blueshifted absorption; we find no evidence of OH emission at blueshifted, systemic, or marginally redshifted (given the limited bandwidth of ALMA) velocities. The fact that blueshifted absorption is observed at velocities far beyond the width of the \cii line is evidence that the high-velocity absorption indeed arises from an outflowing wind.

As described in the main text, we fit the integrated spectrum with a constant continuum flux density and the sum of two OH velocity components, each consisting of a pair of equal-amplitude Gaussian profiles, where one component represents the systemic absorption and one the blueshifted wind absorption. Such a fit is not intended to be unique, but does capture the velocity structure of the data well. In particular, there is also probably additional velocity substructure present in the absorption that cannot be distinguished at the signal-to-noise of our data. There is also probably a more-or-less continuous range of absorbing velocities, without a clear distinction between systemic and wind absorption, as expected because the wind is launched from within the galaxy. The high signal-to-noise spectrum of \cii provides more clarity; it indicates that the emission from gas within the galaxy is confined to velocities $v>-200$\,\kms.  The absorption more blueshifted than this is therefore associated with the wind almost exclusively. For consistency with the literature, we also estimate \vmax, the maximum velocity of the blueshifted line wing, although the determination of this velocity depends somewhat on the signal-to-noise (S/N) of the data, and methods for determining \vmax vary in the literature. We adopt the velocity above which 98\% of the absorption occurs, following directly from our use of Gaussian profiles to fit the spectrum. For SPT2319$-$55, we find $\vmax = -800 \pm 110$\,\kms, consistent with the velocity near which the OH absorption becomes too weak to detect.  Using the best-fit continuum level and Gaussian profiles, we also measure the equivalent width of the systemic and blueshifted components. The equivalent widths have the advantage of being largely independent of gravitational lensing, as both the absorption and continuum are magnified by essentially the same factor. We summarize the properties of the Gaussian profile spectral fitting in Table~\ref{tab:outflowspec}.

\section{Gravitational Lens Modeling}

\subsection{Lens Modeling Methodology}

To investigate the intrinsic source structure of SPT2319$-$55, we must model the effects of gravitational lensing. We use a pixellated source reconstruction code, described in detail in ref. \cite{hezaveh16}. Briefly, the code fits the measured visibilities directly in order to avoid correlated uncertainties inherent to inverted interferometric images. The source plane is represented by a grid of pixels that are mapped to the image (observed) plane by a lensing operator. We assume a singular isothermal ellipsoid mass profile \cite{kormann94} with strength parameterized by the mass interior to $10$\,kpc. This mass profile has been shown to be a good approximation of typical galaxy-scale lensing halos \cite{treu04,koopmans06,koopmans09}.  The lens ellipticity is decomposed along orthogonal axes. The lens position is relative to the ALMA phase center, with the positive x-direction to the west and positive y-direction to the south.

\begin{figure*}[htb]
\centering
\includegraphics[width=\textwidth]{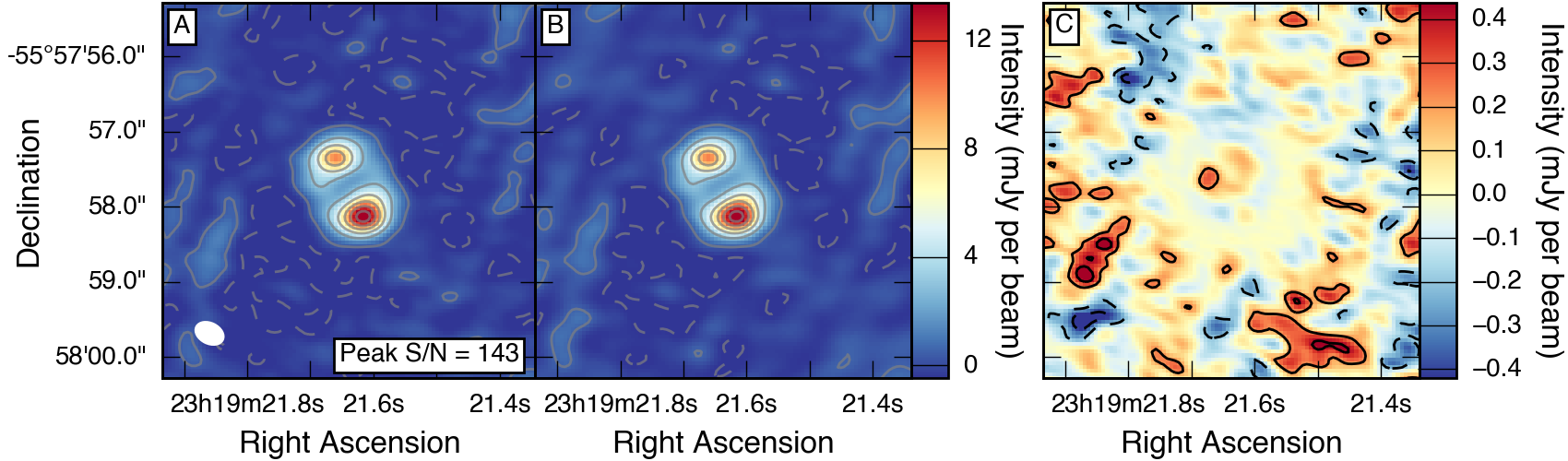}
\caption{
\textbf{Model of the continuum emission of SPT2319$-$55.} 
\textbf{A.} Dirty (not deconvolved) image of the data. An ellipse representing the beam is shown at lower left.
\textbf{B.} Model image.
\textbf{C.} Residuals.
In panels A and B, contours are drawn at $\pm5\sigma$, increasing in steps of 25$\sigma$, where 1$\sigma = 105$\,\uJy\,beam$^{-1}$. In panel C, contours are drawn in steps of $\pm2\sigma$. The structure in the dirty image is due to sidelobes of the synthesized beam, and should be reproduced by the model image in B.
}
\label{fig:outflowcontmodel}
\end{figure*}

The array of source plane pixel values is regularized (effectively smoothed) by a linear gradient on the source, which minimizes the pixel-to-pixel variations in the source plane \cite{warren03,suyu06,hezaveh16}. The strength of the regularization is determined by maximizing the Bayesian evidence of the source-plane reconstruction given a set of fixed lens parameters. Fitting for the strength of the regularization is designed to avoid over- or under-fitting the data. We determine the best-fit lens parameters through a Markov Chain Monte Carlo (MCMC) sampling algorithm, fitting to the continuum (upper sideband) data. We show the continuum model of this source in Figure~\ref{fig:outflowcontmodel}. 

Once the lens parameters are known, we use the best-fit parameters to reconstruct the OH line averaged in velocity as described in the main text.  In principle a joint fit of both the continuum and line data would provide the best possible lens model parameters and source plane reconstructions; unfortunately the sheer number of visibilities and image and source plane pixel values make such a fit computationally infeasible. To reconstruct the OH absorption in SPT2319$-$55, we model three channels which cover almost purely systemic absorption, wind absorption, and line-free continuum emission (i.e., we select a subset of the continuum data with a bandwidth similar to that of the absorption components). We conservatively define the wind and systemic absorption components to include velocities $-700<v<-200$\,\kms and $+300<v<+700$\,\kms relative to the higher-frequency OH transition, respectively. As Figure~1 shows, these velocity ranges are dominated by their respective component; in particular, there is almost no overlap between the selected wind velocity range and the systemic \cii emission. While not ideal -- we have no way to separate strongly blueshifted systemic absorption from mildly blueshifted wind absorption -- this method does at least separate pure wind from pure systemic absorption.

Because the S/N of any given subset of channels will be much lower than the S/N of the continuum emission as a whole due to the different bandwidths involved, the best-fit strength of the source regularization is slightly higher for these velocity ranges than the continuum as a whole. For the purposes of Figure~2 in the main text, we modeled a continuum channel by selecting 500\,\kms of line-free bandwidth from the same sideband as the OH absorption and model this alongside the absorption components. The velocities selected correspond to the range from $-1500<v<-1000$\,\kms in Figure~1. In this way we ensure that each map was reconstructed with the same parameters and from the same amount of data. Using this continuum image, we create continuum-subtracted reconstructions of the molecular wind and systemic absorption shown in Figure~2 of the main text. We have verified that the integrated absorption depths of each component in Figure~1 are consistent with the absorption depths and magnifications implied from our lensing reconstructions.

\subsection{Lens Model Parameters}

\begin{figure*}[htb]%
\centering
\includegraphics[width=\textwidth]{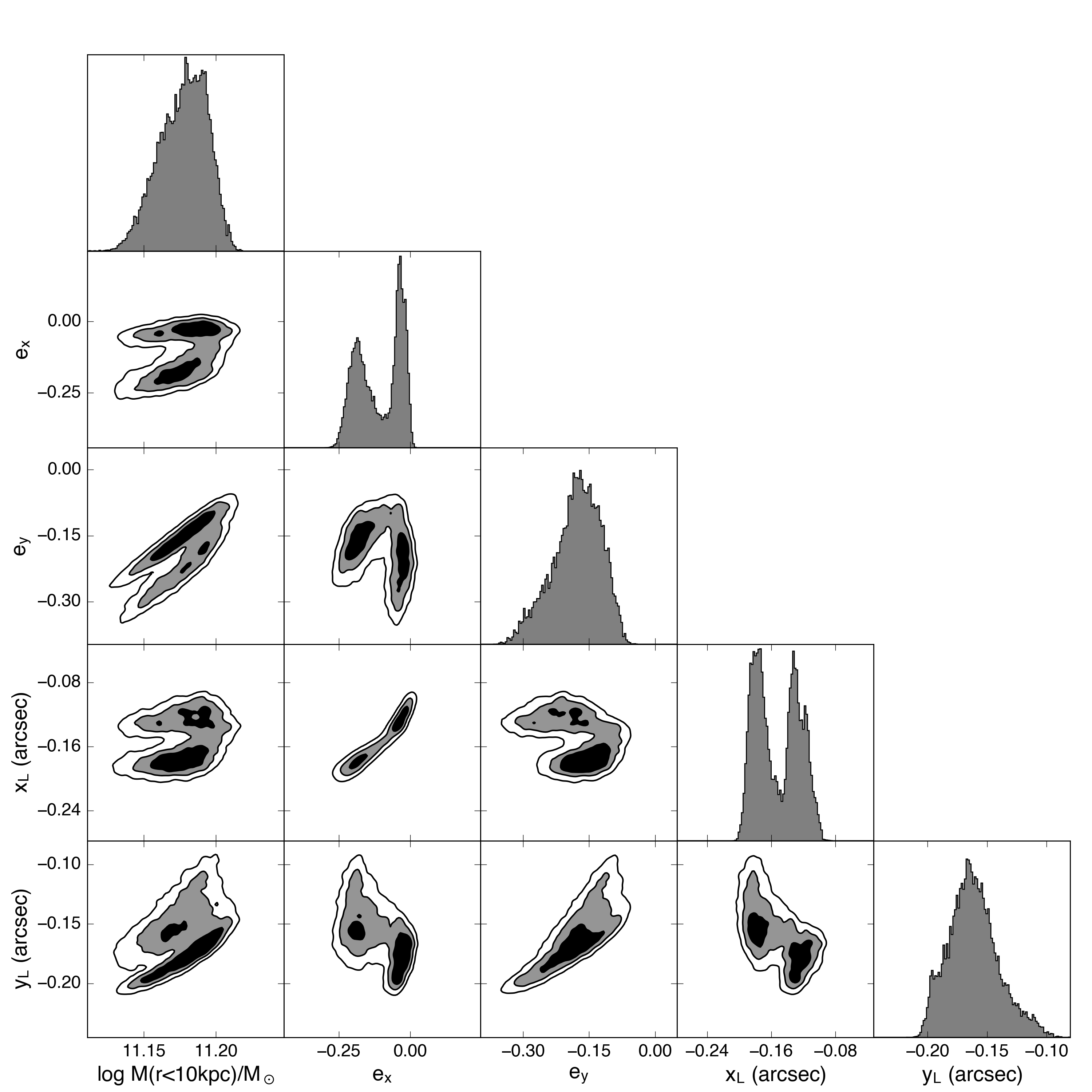}
\caption{
\textbf{Parameter degeneracy plot for the SPT2319$-$55 lens model.} The left-most column indicates the lens mass enclosed within a radius of 10\,kpc. Contours are drawn to enclose 68, 95, and 99\% of the integrated likelihood. One-dimensional marginalized parameter distributions are shown as histograms down the diagonal of the plot.
}
\label{fig:outflowlensdetail}
\end{figure*}

The lens parameter degeneracies are shown in Figure~\ref{fig:outflowlensdetail}. There is a bimodal solution in the lens parameters, with two possible lens positions separated by $\sim0.05$\arc and correspondingly different lens ellipticities. This lens position separation is smaller than the astrometric uncertainty in other available data, so we have no external information to break this bimodality. Additionally, a parameter degeneracy between the lens mass and ellipticity is apparent due to the particular implementation of the lensing deflections used by this reconstruction code, which includes a factor proportional to the lens ellipticity in the conversion between lens mass and deflection strength. We have verified that the source-plane reconstructions at both maxima in the likelihood are very similar in both the continuum and the systemic and blueshifted absorption components, and the differences have no appreciable impact on our further analysis. 

\begin{figure}[htb]
\centering
\includegraphics[width=0.4\textwidth]{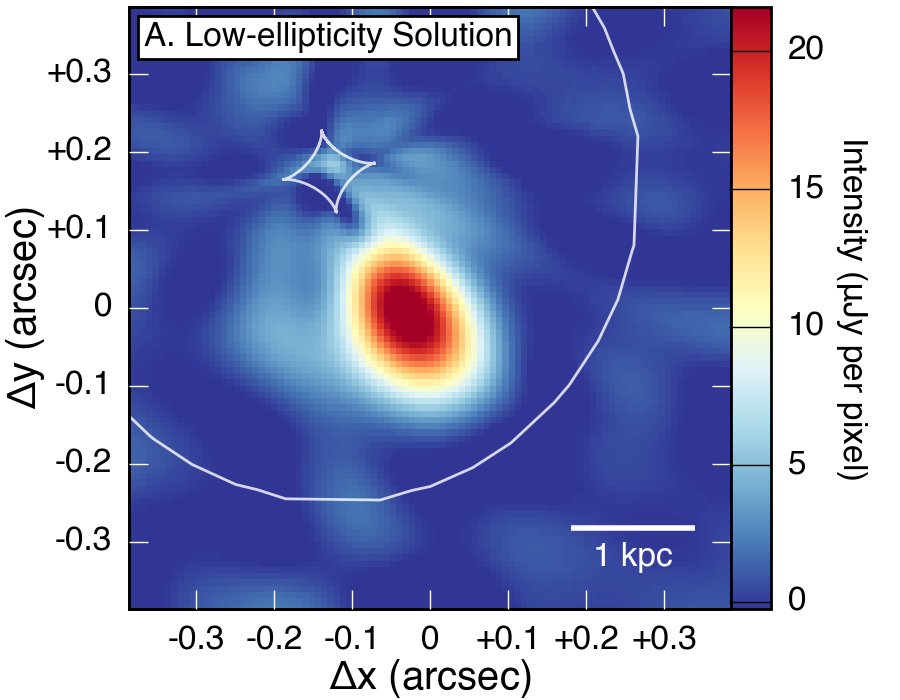}
\includegraphics[width=0.4\textwidth]{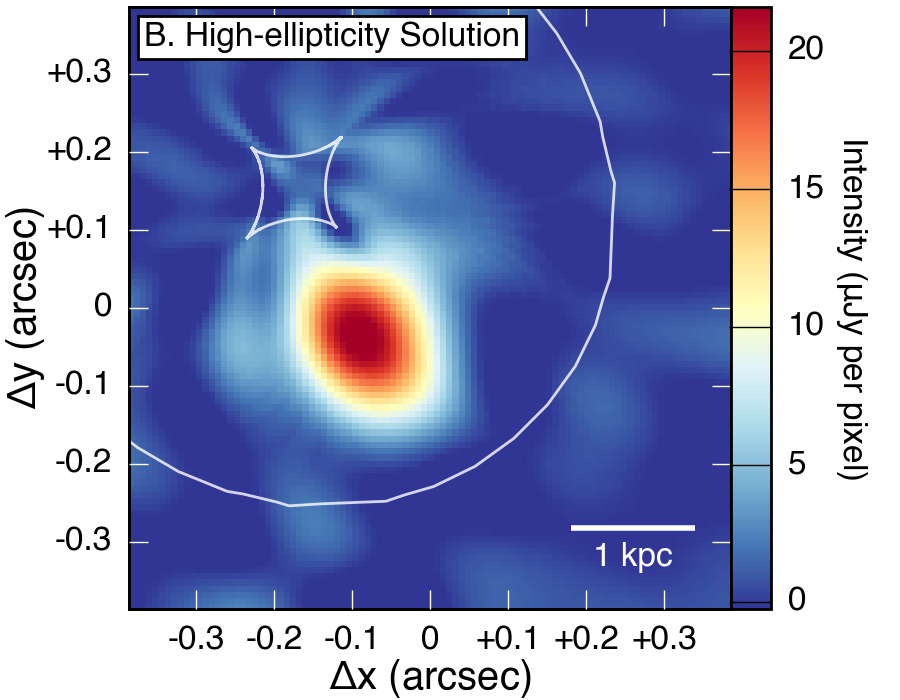}
\caption{
\textbf{Source-plane reconstructions of SPT2319$-$55 for each of the two bimodal lens solutions in Figure~\ref{fig:outflowlensdetail}.} 
\textbf{A.} The solution with lower lens ellipticity. \textbf{B.} The solution with higher lens ellipticity. Little qualitative or quantitative differences are seen; we adopt the solution shown in panel A. The lensing caustics are shown in white.
}
\label{fig:twolenssols}
\end{figure}

The source-plane reconstructions of the continuum at each of the maxima in the likelihood are shown in Figure~\ref{fig:twolenssols}. For the main body of this work, we show models assuming the solution with slightly lower lens ellipticity and position relative to the ALMA phase center $x_L = -0.13$\arc, $y_L = -0.18$\arc.

In both solutions, the source is dominated by a single large emitting region $\sim$1.2\,kpc in diameter (FWHM) -- we have no evidence for small-scale substructure or clumpiness in the dust emission.  Whether this smooth structure would also be recovered by data with $\sim$10$\times$ higher resolution, as ALMA is capable of delivering, is difficult to predict (see discussion in \cite{spilker16}). This scenario is no different than that of unlensed sources, which also cannot resolve structure below the resolution limit. Even when the resolution is high, discerning the presence or absence of genuine ``clumps'' of emission is difficult. Even images which appear to have clumpy structure can in fact be consistent with smooth emission due to limited signal-to-noise and low contrast between clumps and the extended emission \cite{hodge16}. Our data yield a peak signal-to-noise $\gtrsim4\times$ higher than even the most significantly detected case considered by ref. \cite{hodge16}, which should lessen these effects.

This source was also modeled by ref. \cite{spilker16} using ALMA data taken in 2012 of lower spatial resolution and peak signal-to-noise than the data now available. That model indicated that the bulk of the dust emission was located northeast of the lens position instead of southwest as the current models show. The reason for the discrepancy is the symmetry between the source and lens positions in doubly-imaged sources. For a doubly-imaged source, the lens and source positions can be transposed without significantly affecting the observed emission. The degeneracy can in principle be broken by constraining the location of the lens through other means (i.e., optical or near-infrared imaging), but in this case the lens position shifts by $\sim0.25$\arc along the direction of the lens major axis; in any case, such a small shift is comparable to the absolute astrometric registration between our best-available near-infrared imaging and ALMA. The degeneracy can also be broken using the fact that the background source is not point-like and so does not produce a simple double image configuration. This requires the source to be sufficiently spatially resolved to detect the subtle difference between the solutions, as in the present higher-resolution data but (apparently) not the 2012 data.

\subsection{Lens Model Tests}

\subsubsection{Effective Source-plane Resolution}

The pixellated reconstruction technique we employ affords substantial freedom in the background source morphology in order to reproduce the data. While the data in Figure~1 clearly resolve SPT2319$-$55, it is not straightforward to translate the spatial resolution of the data to an effective resolution in the source plane. Similarly, it is possible that the lensing reconstruction procedure could introduce artifacts that lead to the clumpy structure seen in the molecular outflow in Figure~2. To test the resolution of the source-plane reconstructions we derive in this work, we ran a series of reconstructions of mock data.

To test the reconstruction of the continuum emission (Figure~2A), we insert a pointlike background source into the source plane, lens this source, and sample the Fourier transform of the resulting lensed image at the same $uv$ coordinates and with the same noise properties as the real data. The intrinsic flux density of the source is set such that the apparent flux density is the same as the observed, $\approx$52\,mJy, ensuring the same signal-to-noise ratio in the mock data as the real observations. We then reconstruct the source plane in the same manner as the real data, again fitting for the strength of the source-plane regularization. Finally, we measure the size of the reconstructed source by fitting a simple two-dimensional Gaussian profile to the source-plane image. 

\begin{figure*}[htb]
\centering
\includegraphics[width=\textwidth]{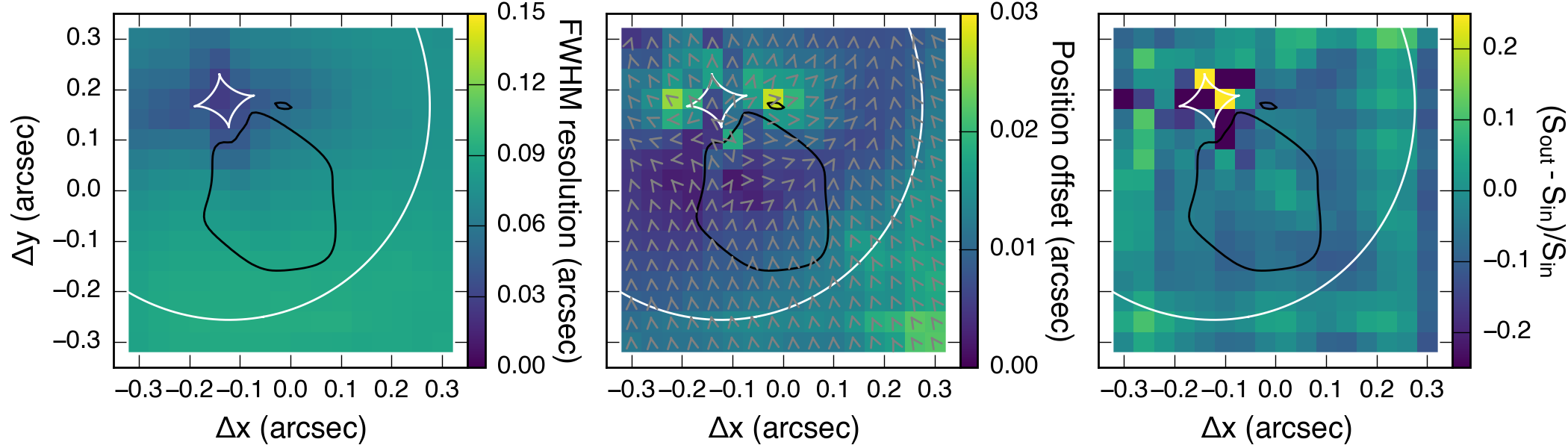}
\includegraphics[width=\textwidth]{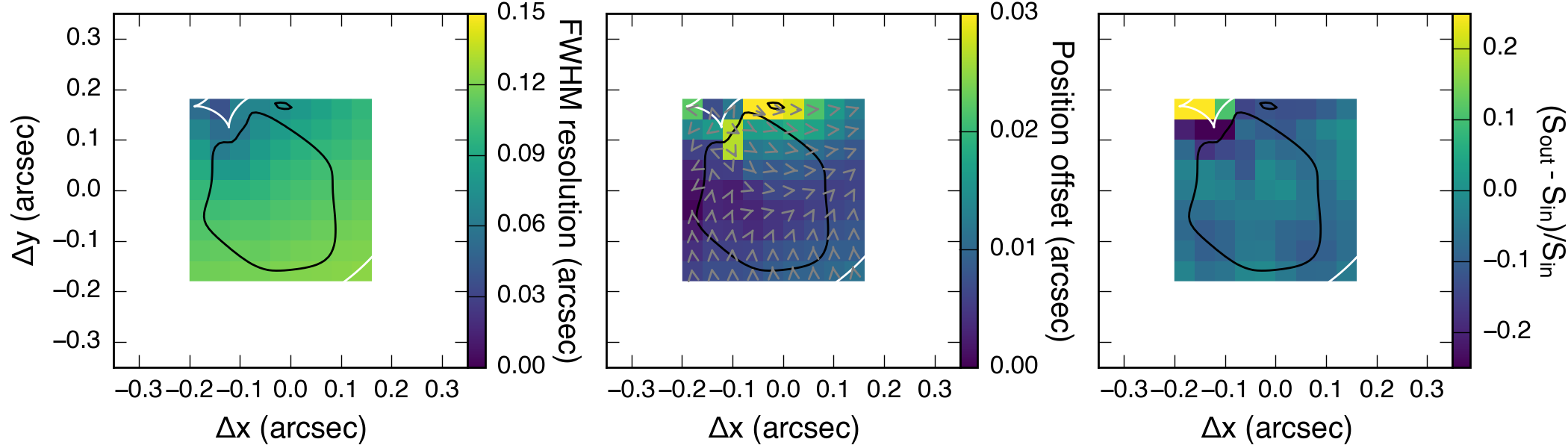}
\caption{
\textbf{Testing the resolution, positional offset, and flux recovery of the lens models.} The top row shows the results of testing the continuum reconstruction; the bottom row shows the results for the absorption components. The left column shows the circularized full-width-at-half-maximum (FWHM) resolution.  The center column shows the difference between input and recovered source positions with arrows denoting the direction of the shifts (the actual shifts are smaller than the arrows). The right column shows the fractional flux recovered compared to the input fluxes. In all panels, the lensing caustics are shown in white, and the black contour indicates the outermost contour drawn in Figure~2, corresponding to signal-to-noise of 8 in the continuum reconstruction.
}
\label{fig:srcresolution}
\end{figure*}

We perform this procedure for background point sources tiled across the source plane, effectively measuring the point-spread function as a function of background source position. Manual inspection of the fits verifies that Gaussians are generally good representations of the reconstructed sources, and the reconstructions show no artificially-introduced clumpy structure. Figure~\ref{fig:srcresolution} shows the resulting map of the FWHM resolution of the reconstructions, as well as the difference between the positions of the input and recovered sources. From this figure we conclude that the lensing procedure does not introduce significant positional shifts in the vicinity of the continuum emission of SPT2319$-$55 ($\lesssim$0.01\arc, or 60\,pc), and that the effective source-plane resolution in this region is $\approx$0.08\arc.  This implies $\approx$10--15 independent resolution elements across the extent of the source. For comparison, the original (observed, apparent) data span $\approx$15--20 independent beams at the resolution of the ALMA data.

Testing the effective resolution of the systemic and blueshifted absorption components is less straightforward because the detection of absorption requires the presence of continuum emission, and the continuum strength, absorption strength, and effective resolution all vary across the source plane. To keep this test simple yet meaningful, we simply repeat the same procedure as for the continuum, but instead of inserting a source to ensure a constant apparent flux density, we instead insert sources of constant \textit{intrinsic} flux density. We choose this intrinsic flux density to be approximately a factor of 3 weaker than either of the two main clumps seen in the outflow reconstruction in Figure~2, in order to test our sensitivity to weaker absorption components than those actually observed. Additionally, we now fix the source-plane regularization strength to the best-fit value from the continuum reconstruction, as was done for the real data. The results of this test are also shown in Figure~\ref{fig:srcresolution}, performed over a smaller region in the source plane corresponding to the general location of the continuum emission of SPT2319$-$55. 

For the absorption components, we again see no evidence for significant positional shifts or artificially-introduced clumpy structure. The source-plane resolution in the vicinity of the SPT2319$-$55 emission is $\approx$0.1\arc, unsurprisingly somewhat worse than for the higher signal-to-noise continuum reconstruction. The absorption maps in Figure~2 are spatially resolved, with $\approx$10 independent resolution elements across the source. The two main absorption clumps in the molecular outflow are separated by $\sim$2.5 FWHM resolution elements; their separation is significant. Additionally, the input flux density of the absorption components is recovered to $\sim$10\% in the vicinity of the observed absorption. From this test we conclude that the reconstruction procedure does not introduce artificial clumpy structures, accurately recovers the intrinsic flux density of the source, that our data are sensitive to absorption components at least $\sim3\times$ weaker than observed, and that both the continuum and absorption maps are spatially resolved into $\gtrsim$10 resolution elements.

\subsubsection{Effects of Source-plane Regularization}

We also explore the extent to which the source-plane reconstruction depends on the regularization parameter $\lambda$, which effectively acts to smooth the source plane emission \cite{suyu06,hezaveh16}. We employ a gradient-type regularization. We fit for the value of $\lambda$ by maximizing the evidence given a set of lens parameters to avoid over- or under-fitting the data.  The reconstructions we use, therefore, formally correspond to the ``optimum'' maps we can construct using the current data. While stronger regularization can produce source-plane maps that look better by eye and have very high inferred signal-to-noise, these maps are not statistically justified; they are very precise, but inaccurate \cite{suyu06}.

Nevertheless, it is worthwhile to explore the differences in source-plane structure induced by different levels of regularization. In Figure~\ref{fig:outflowregularization}, we show reconstructions with a wide variety of regularization relative to the best-fit value \Lz. For reference, changing the lens parameters typically changes the best-fit value of $\lambda$ by $\lesssim20$\%, so the strengths of regularization shown in Figure~\ref{fig:outflowregularization} are exaggerated compared to the true uncertainty in $\lambda$. Each panel is shown with the same scaling relative to the fiducial (center) image to accentuate the differences in the other reconstructions. We find very similar structures across all values of the source regularization, and the total flux of the source changes by $<10$\%. As the strength of the regularization increases, however, we do see that some of the ``noise'' in the images is suppressed, while the source becomes slightly more extended and diffuse by $\sim$20\%. The degree of qualitative similarity and small quantitative differences even with large variations of $\lambda$ is evidence that the source-plane structure observed is accurate. 

\subsubsection{Consistency of Continuum Reconstructions}

\begin{figure*}[htb]%
\centering
\includegraphics[width=\textwidth]{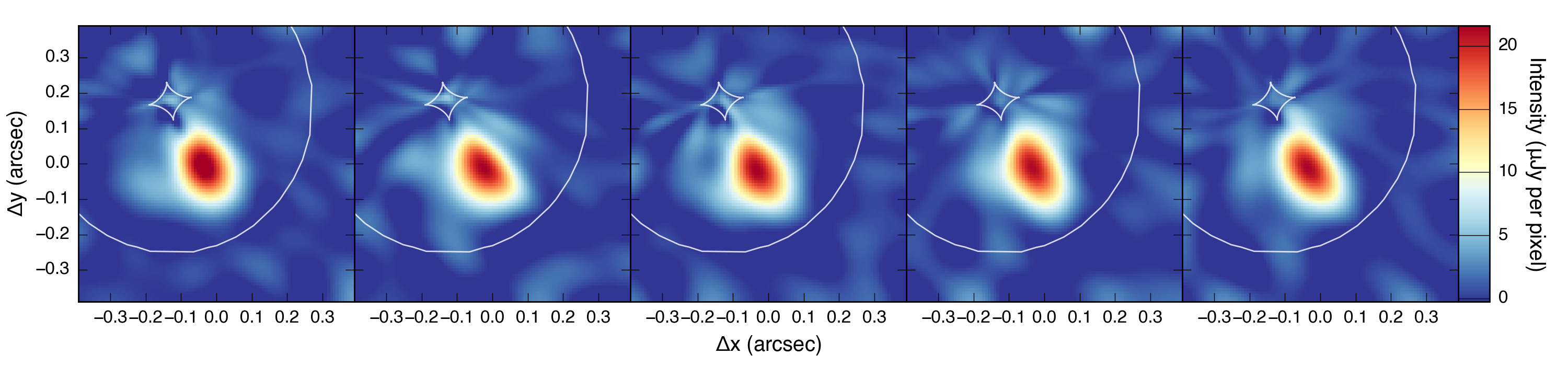}
\caption{
\textbf{Testing the consistency of the continuum reconstruction.} Each panel shows the lensing reconstruction of an independent 500\,\kms channel, taken from the line-free ALMA upper sideband data. Each is shown on the same color scale, and the white lines show the lensing caustics.
}
\label{fig:contrealization}
\end{figure*}

\begin{figure*}[htb]%
\centering
\includegraphics[width=\textwidth]{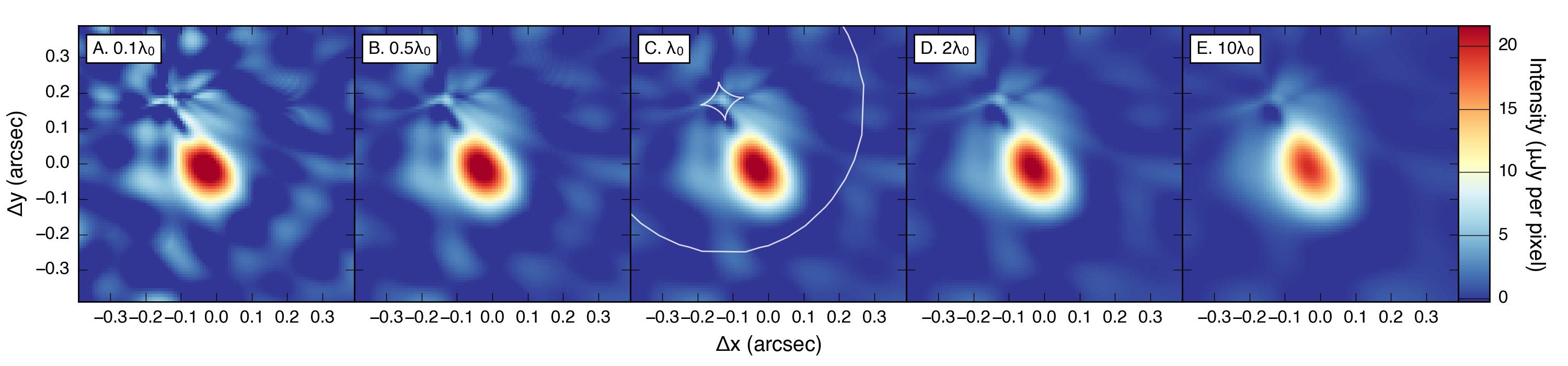}
\caption[Effects of the source regularization strength]
{
\textbf{Testing the effects of the strength of source regularization.} Panels A--E increase the strength of the regularization compared to the best-fit value from the real data $\lambda_0$ (panel C). Each image is displayed on the same linear color scale, set by the middle panel, to accentuate the differences with the other panels. In the middle (fiducial) panel, we also plot the lensing caustics.
}
\label{fig:outflowregularization}
\end{figure*}

The reconstruction of the continuum emission was created by selecting data in a 500\,\kms channel taken from the lower sideband, which also contains the OH absorption. Here we briefly test whether the continuum morphology seen in that reconstruction is also found in other 500\,\kms realizations of the continuum emission. While the continuum bin shown in Figure~2 is the only line-free bin of that width available in the lower (OH-containing) sideband (Figure~1), we can also construct an additional five independent such bins using the data from the ALMA upper sideband, which spans $\approx$2600\,\kms centered at observed frequency 411.82\,GHz. The result is shown in Figure~\ref{fig:contrealization}. While five reconstructions is too few to measure a meaningful pixel-by-pixel standard deviation in the maps (and therefore an independent estimate of the signal-to-noise ratio of the reconstruction), there is little qualitative or quantitative variation in the regions with significant continuum emission shown in Figure~\ref{fig:contrealization}.

\section{Mass Outflow Rate Determination}

We consider a simple geometric model for the molecular outflow in SPT2319$-$55; we do not attempt to match, for example, the velocity profiles of the systemic and wind absorption or the spatial distribution of the outflowing material. The mass outflow rates derived in this section should be considered highly uncertain. The primary uncertainties arise from the unknown optical depth of the OH 119\,\um doublet, the OH abundance relative to H$_2$, and the detailed geometry of the wind. Geometrical uncertainties include both the unknown radial separation between the absorbing gas components and the continuum emitting region as well as  any additional outflowing gas along lines of sight that do not intersect the continuum emitting region.

We estimate the minimum column density of OH, $N_{\rm{OH}}$, by assuming optically thin absorption. The total column density can be written 
\begin{equation} \label{eq:Ntot}
N_{\rm{OH}} = \frac{8 \pi Q_{\rm{rot}}(T_{\rm{ex}})}{\lambda^3 g_u A_{ul}} \frac{\exp(E_l/T_{\rm{ex}})} { 1 - \exp\left(\frac{-h \nu}{k T_{\rm{ex}}}\right)} \int \tau d\rm{v},
\end{equation}
where $\lambda$ is the wavelength, $h$ and $k$ are the Planck and Boltzmann constants, $A_{\rm{ul}}$ the Einstein A coefficient of the transition, $g_u$ the degeneracy of the upper energy level, $E_l$ the lower energy level in temperature units, $Q_{\rm{rot}}$ the rotational partition function evaluated at excitation temperature $T_{\rm{ex}}$, and $\int \tau d\rm{v}$ the integrated optical depth of the absorption line \cite{mangum15}, which we measure from our spectral fitting procedure. For the ground-state OH 119\,\um transition, $E_l = 0$\,K. In the optically thin regime, the integrated optical depth is simply the equivalent width of the wind component. The values of the energy and degeneracy levels, Einstein A coefficient, and tabulated values of the partition function are available from the NASA JPL and/or LAMDA spectroscopic databases \cite{muller01,muller05,schoier05}. We adopt an OH abundance relative to H$_2$ of $5 \times 10^{-6}$ \cite{goicoechea02} to convert the OH column density to a total molecular gas column density, as commonly assumed in the literature (though see also ref. \cite{richings18}, who find OH abundances relative to H$_2$ up to a factor of $\sim$4 higher in theoretical outflow models which include a chemical reaction network, because OH is also present in atomic gas in these models). If the metallicity of SPT2319-55 is sub-solar, leading to lower OH abundance, our estimates of the H$_2$ column density and the mass outflow rate would increase proportionally. Because the excitation of the higher-lying energy levels is dominated by pumping by FIR continuum photons (at 35, 53, 65, and 84\,\um, among others) instead of collisions with H$_2$, the OH level populations are mostly dominated by the FIR radiation field generated by heated dust \cite{gonzalezalfonso14,gonzalezalfonso17}. We assume $T_{\rm{ex}} = \Tdust = 100$\,K as found for high-velocity OH winds in the literature \cite{gonzalezalfonso14}; the column density varies by less than a factor of 2 for $50<T_{\rm{ex}}<160$\,K. 

This column density can be converted to an estimated mass outflow rate by further assuming an outflow geometry. We used the time-averaged thin-shell approximation of \cite{rupke05} to calculate the molecular mass outflow rate \Mdot and outflow mass \Mout for an outflow at radius $R$ and velocity $v$:
\begin{align} \label{eq:mdot}
\begin{split}
\Mdot &= 4 \pi R^2 \: f_{\rm{cov}} \: \mu m_{\rm{H}} \: N_{\rm{H}} \: v /  R , \\
\Mout &= \Mdot \: R / v,
\end{split} \end{align}
with $f_{\rm{cov}}$ the covering factor of the wind material,$\mu=1.36$ the mean mass per hydrogen nucleus (including the cosmological helium abundance), m$_{\mathrm{H}}$ the mass of a hydrogen nucleus, and N$_\mathrm{H}$ the column density of hydrogen.
This is a conservative choice relative to other geometries, and is generally appropriate for comparison with outflow models \cite{gonzalezalfonso17}. We assume that the covering factor along the line of sight we observe, $\sim$0.8, is typical of the wind as a whole, that the wind material is at a typical distance comparable to the galaxy continuum-emitting size, $\sim$600\,pc, and use the mean wind velocity 440\,\kms. This too is expected to be a conservative assumption, as the dust continuum-emitting region is generally found to be more compact than the galaxy size measured for other tracers \cite{spilker15,ivison13,chen17}.

\begin{figure*}[htb]
\centering
\includegraphics[width=0.5\textwidth]{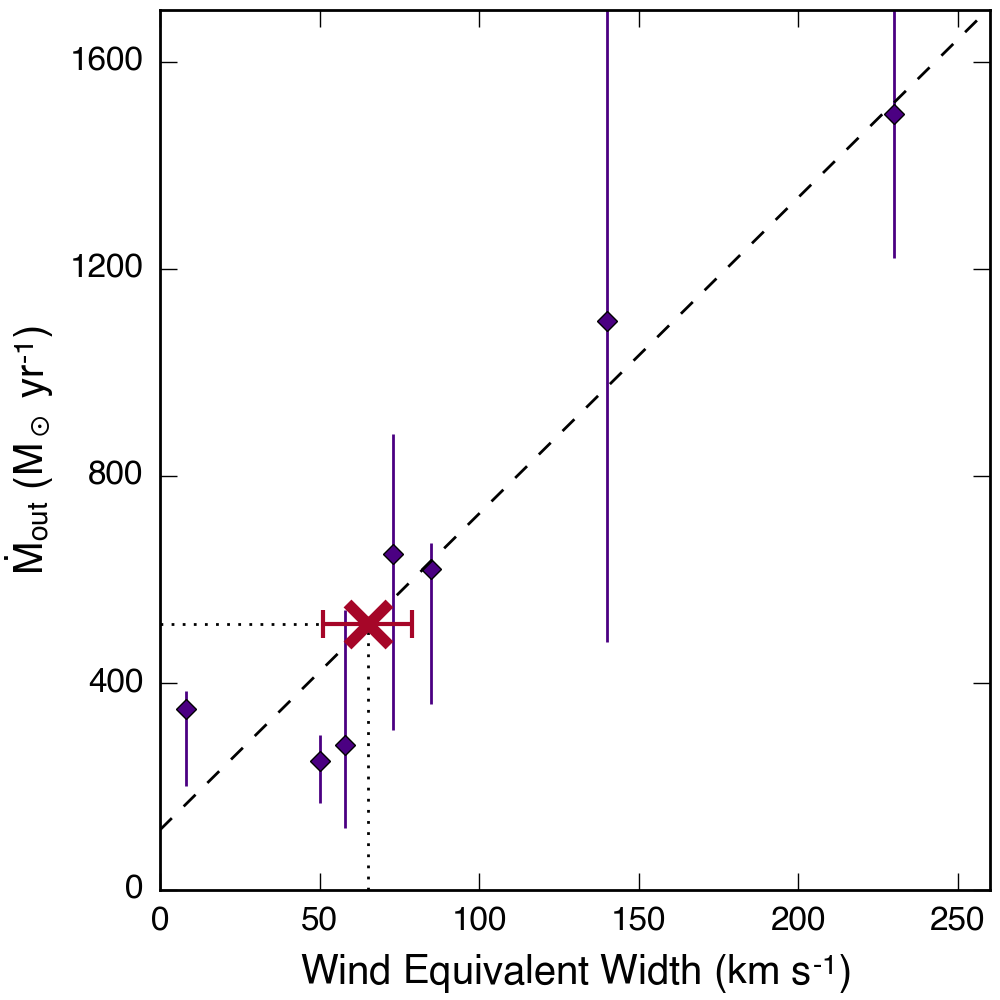}
\caption{
\textbf{The outflow rates of low-redshift dusty galaxies as a function of the OH 119\,\um equivalent width \cite{gonzalezalfonso17}.}
The equivalent width has been integrated over the blueshifted velocities corresponding to the outflowing material. A linear fit to these points is plotted with a dashed line. If SPT2319$-$55 also follows this correlation, marked as an $\times$, the implied mass outflow rate is $\sim$510\,\Msol\,yr$^{-1}$.
}
\label{fig:weqmdot}
\end{figure*}

Under these assumptions, we find a minimum mass outflow rate $\gtrsim$60\,\Msol\,yr$^{-1}$. The fact that absorption in higher-lying OH transitions is commonly observed in low-redshift dusty galaxies (in particular the cross-ladder transitions at 35, 53, and 79\,\um) has been taken as evidence that the 119\,\um doublet must be very optically thick \cite{gonzalezalfonso14,gonzalezalfonso17}. The optically-thin minimum outflow rate thus severely underestimates the true outflow rate. However, a reasonable correlation appears to exist between the  mass outflow rate and the OH 119\,\um equivalent width integrated over the blueshifted line wing associated with the outflow. Figure~\ref{fig:weqmdot} shows the relationship between the outflow rates determined through modeling of multiple OH transitions in local dusty galaxies as a function of the OH 119\,\um equivalent width \cite{gonzalezalfonso17}, integrated over the velocity ranges associated with emission from outflowing CO \cite{cicone14}. If SPT2319$-$55 also follows this correlation, it may imply that the small-scale structure or clumpiness of molecular winds is fairly similar in galaxies, allowing the outflow rate to be estimated despite the high optical depths of the 119\,\um transition. While perhaps surprising, a similar argument applies to CO itself, which is nearly always optically thick yet still provides a reasonably faithful measure of molecular gas masses. If SPT2319$-$55 also follows this correlation, the implied outflow rate is $\sim$510\,\Msol\,yr$^{-1}$ and the total outflow mass is $\sim7\times10^8$\,\Msol; in other words, the assumption of optically thin absorption underestimates the outflow rate by about an order of magnitude. Given the typical optical depth of OH 119\,\um, such a large difference is unsurprising. Again, if the metallicity of SPT2319-55 is significantly sub-solar, the mass outflow rate would increase (and the true outflow rate of SPT2319-55 would not lie on the empirical correlation defined by the low-redshift objects). We stress again that both estimates carry large systematic uncertainties of at least a factor of 2, possibly rising to a factor of 10 if the assumed geometry is very different from reality.

\clearpage

\clearpage

\begin{deluxetable}{lr}
\tabletypesize{\scriptsize}
\tablewidth{0pt}
\tablecolumns{2}
\tablecaption{Basic Properties of SPT2319$-$55\label{tab:outflowspt2319data}. The effective radius \reff is measured from the 119\,\um continuum emission.}
\tablehead{\colhead{Parameter} & \colhead{Value}}
\startdata
Full Source Name    & SPT-S J231921-5557.9 \\
Right Ascension     & 23h 19m 21.67s \\
Declination         & -55d 57m 57.8s \\
$z_{\rm{lens}}$     & 0.91 \\
$z_{\rm{source}}$   & 5.2943 \\
\lir (8--1000\,\um) & (7.9 $\pm$ 3.0) $\times10^{12}$ $(\mu/5.8)^{-1}$ \Lsol \\
\lfir (40--120\,\um)& (4.3 $\pm$ 0.8)$\times10^{12}$ $(\mu/5.8)^{-1}$ \Lsol \\
SFR                 & 790 $\pm$ 300 $(\mu/5.8)^{-1}$ \Msol\,yr$^{-1}$ \\
\reff$^b$           & 0.6 $\pm$ 0.1 kpc \\
\Mgas               & (1.2 $\pm$ 0.2) $\times10^{10}$ $(\mu/5.8)^{-1}$ \Msol \\
\enddata
\end{deluxetable}

\begin{deluxetable}{lr}
\tabletypesize{\scriptsize}
\tablewidth{0pt}
\tablecolumns{2}
\tablecaption{Spectrum Fitting Results\label{tab:outflowspec}. Outflow component velocity is relative to the higher-frequency line of the OH doublet, as in Figure~1. Parameters marked with asterisks are derived, not fitted. Equivalent widths are reported for a single line of the OH doublet transition (i.e., they should be multiplied by a factor of two for the total absorption strength).}
\tablehead{\colhead{Parameter} & \colhead{Value}}
\startdata
Continuum level & $52.1 \pm 0.5$ mJy \\
\multicolumn{2}{c}{Systemic Component}\\
Absorption amplitude & $-7.8 \pm 1.2$ mJy \\
FWHM & $330 \pm 80$ \kms \\
Equivalent Width* & $52 \pm 15$\,\kms\\
\multicolumn{2}{c}{Outflow Component}\\
Absorption amplitude & $-7.0 \pm 1.2$ mJy \\
Center velocity & $-440 \pm 50$ \kms \\
FWHM & $450 \pm 60$ \kms \\
Maximum velocity* & $-800 \pm 110$\,\kms \\
Equivalent Width* & $64 \pm 14$\,\kms
\enddata
\end{deluxetable}

\end{document}